# Understanding Perceptions and Attitudes in Breast Cancer Discussions on Twitter

François Modave[a*], Yunpeng Zhao[b*], Janice Krieger[c], Zhe He[d], Yi Guo[b], Jinhai Huo[e], Mattia Prosperi[f], Jiang Bian[b]

[a] *Department of Medicine, Center for Health Outcomes and Informatics Research Loyola University Chicago, Maywood, IL, USA,*
[b] *Department of Health Outcomes and Biomedical Informatics, University of Florida, Gainesville, FL, USA,*
[c] *Department of Advertising, University of Florida, Gainesville, FL, USA,*
[d] *School of Information, Florida State University, Tallahassee, Florida, USA.*
[e] *Department of Health Services Research, Management and Policy, University of Florida, Gainesville, Florida, USA*
[f] *Department of Epidemiology, University of Florida, Gainesville, Florida, USA*
[*] *Equal contribution, co-first*

**Abstract**

*Among American women, the rate of breast cancer is only second to lung cancer. An estimated 12.4% women will develop breast cancer over the course of their lifetime. The widespread use of social media across the socio-economic spectrum offers unparalleled ways to facilitate information sharing, in particular as it pertains to health. Social media is also used by many healthcare stakeholders, ranging from government agencies to healthcare industry, to disseminate health information and to engage patients. The purpose of this study is to investigate people's perceptions and attitudes relate to breast cancer, especially those that are related to physical activities, on Twitter. To achieve this, we first identified and collected tweets related to breast cancer; and then used topic modeling and sentiment analysis techniques to understand discussion themes and quantify Twitter users' perceptions and emotions w.r.t. breast cancer to answer 5 research questions.*

## Introduction

A report from the National Cancer Institute (NCI) indicates that one in eight women will develop breast cancer during the course of her lifetime [1]. An estimated 266,120 new cases of invasive breast cancers, and 63,960 non invasive breast cancers will be diagnosed in 2018 in the U.S.. Among American women, breast cancer remains the second most diagnosed cancer, just behind lung cancer [2]. Nevertheless, a recent study shows that physically active women have a lower risk of breast cancer than inactive women [3]. Further, for breast cancer survivors, physical activities (PAs) have benefits on their mental health, physical conditions, and movement, which ultimatedly improve patients' quality of life [4].

Access to care, access to adequate health information, and health literacy largely remain to be significant issues especially in disenfranchised populations and minorities [5], despite improvements following the implementation of the Affordable Care Act (ACA). However, the widespread availability and uptake of Internet technologies across the socio-economic spectrum has the potential to facilitate health information sharing. Specifically, nearly 9 in 10 Americans have access to high-speed Internet and 7 in 10 use at least one social media platform [6]. Not only are private citizens widely using social media resources, but the various healthcare stakeholders, e.g. industry, governmental agencies, healthcare professionals, have been increasingly using these online platforms to disseminate health information, engage patients, and recruit for clinical trials. Social media platforms provide unique sources of essentially endless data stream, voluntarily shared by their users. These user-generated data provide unique insights into public health; and if mined appropriately, these data are invaluable for understanding various social and health issues.

Twitter is a particularly relevant and effective data source to understand how users' perceptions and attitudes towards health-related issues change over time. In our previous work, we used Lynch syndrome as a case study to show that Twitter can be used effectively to explore discussion topics, and how promotional information can impact laypeople's discussions [7]. In this paper, we describe our Twitter analysis pipeline, as it pertains to users' general perceptions and attitudes towards breast cancer and more specifically whether and how PAs were discussed in these breast cancer-related tweets. We specifically addressed the following five research questions (RQs):

- RQ1: How do people's attitudes (i.e., emotions) towards breast cancer change over time?
- RQ2: How do people's attitudes towards breast cancer differ across geospatial regions?
- RQ3: Can we identify latent topics/themes and topic trends in breast cancer-related tweets?
- RQ4: How does promotional information impact laypeople discussion themes over time?
- RQ5: How physical activities were discussed in laypeople's breast cancer-related tweets?

## Methods

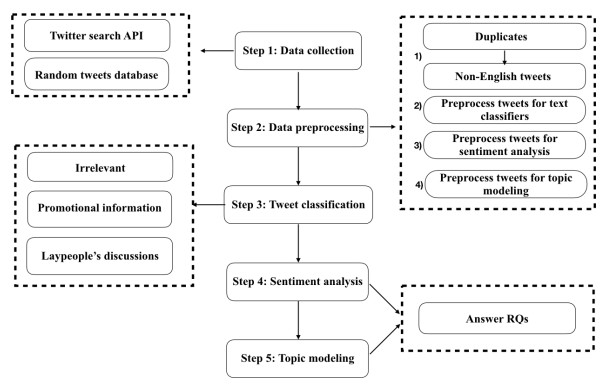

*Figure 1– The workflow of our Twitter analysis pipeline.*

Our approach started with collecting tweets that are relevant to breast cancer discussions. We then classified these tweets into three groups (i.e., irrelevant, promotional information, and laypeople's discussions), assessed laypeople's attitudes (i.e., emotional states) using sentiment analysis, and explored latent themes using a topic modeling approach on both promotional and laypeople's tweets. Finally, based on the sentiment analysis and topic modeling results, we addressed the 5 RQs. Figure 1 illustrates our analysis workflow in 5 steps.

**Step 1: Data collection.** The data used in this study were from two different sources: 1) we collected breast cancer tweets from May 27, 2018 to October 13, 2018 (139 days) using a Twitter crawler based on a set of keywords related to breast cancer (e.g., "*breast cancer*" and "*#BreastCancerFighter*"). The keywords were generated through a snowball sampling process, where we started with seed keywords (e.g., "*#BreastScreening*" and "*#lumpectomy*"), then searched on Twitter with these keywords to retrieve a sample of tweets, evluated the relevance of each tweet, and identified new relevant keywords. We did this process iterately unitl no new keywords were found; and 2) we used the keywords developed above to identify related tweets on a database of public random tweets, which we collected using the Twitter steaming application programming interface (API) from January 1, 2013 to December 30, 2017.

**Step 2: Data preprocessing.** We preprocessed the collected data to eliminate tweets that were 1) duplicates across the two sources, or 2) not in English.

To develop tweet classification models (see Step 3), we preprocessed the tweets following the steps used by Glove [8]: 1) replaced hyperlinks (e.g., "http://t.co/xQgeMny5") with "<url>", 2) replaced mentions (e.g., @Channel9") with "<user>", 3) replaced hashtags (e.g.,"#breastcancer") with "<hastag>", and 4) all emojis were replaced with "<emojies>".

For sentiment analysis, we preprocessed the data with the following steps: 1) removed hyperlinks, 2) removed mentions, 3) converted hashtags into original English words (e.g., converted "#breastcancer" to "breastcancer"), 3) removed all emojies, and 4) geocoded each tweets with a geocoding tool we developed previously [9]. For topic modeling, we lemmatized each word and removed stop words (e.g., "this" and "is").

**Step 3: Tweet classification.** Even though a tweet contains keywords related to breast cancer, the tweet may not be relevant to the breast cancer discussion (e.g., "*I am going to write #fiction about a 40+ year old mom with #oneboob who finds love and life's meanings who wants to read*"). Thus, we developed a two-step process with two classification models to categorize the massive amount of tweets into 3 groups (i.e., irrelevant, promotional, and laypeople's discussions).

We first annotated 1,774 tweets randomly selected from the overall dataset to create a training set. We then experimented with two different deep learning algorithms: convolutional neural networks (CNN) and long short-term memory (LSTM). We implemented both the CNN- and LSTM-based models in Keras on top of the Tensorflow framework.

A common strategy for building deep learning sentence classifiers is to use word embeddings to transform raw texts into vectors of real numbers as features. Thus, we initialized the embedding layer with a pretrained Twitter word embeddings (i.e., 100 dimension) from GloVe. The same feature matrics were used in the two-step classification process: one that categorized tweets into relevant vs. irrelevant, and another one that further categorized the relevant tweets into promotional information vs. laypeople's discussions. Models whith the best performance were adopted to classify the rest of the tweets.

**Step 4: Sentiment analysis**. The Linguistic Inquiry and Word Count (LIWC) is a text analysis tool, which can assess indivdiuals' attitudes through counting the percentage of emotional words used in a given text. LIWC has been used widely and its validity and reliability were validated [10]. We used LIWC on all laypeople's discussions to assess their attitudes/emotions in 5 aspects (i.e., positive emotion, negative emotion, anxiety, anger, and sadness).

**Step 5: Topic modeling**. Topic modeling is a statistical, unsupervised appraoch that can discover abstract themes in a collection of documents. We used the Biterm algorithm to find the main topics in all relevant tweets (i.e., combined both promotional information and laypeople's discussions). Different from conventional topic modeling approaches (e.g., latent Dirichlet allocation) that are based on word-document co-occurrences, Biterm learns topics by modeling word-word co-occurrences patterns, which performs better on short texts [11]. Although topic modeling is a unsupervised method, the number of topics is a parameter that needs to be determined *a priori*. Based on our previous work [7], to capture as many topics as possible, we set the number of topics as 100, visualized the topics in wordclouds, and then manually evaluated each topic's quality and merged topics with similar themes.

To answer our RQs, we also need to know the topic of each tweet. The Biterm model can infer the topic of a given tweet and return a list of topics and associated topic probabilities. We extracted the topic with the highest probability for each tweet.

## Results
### Data collection
Our data came from two different sources as shown in Table 1. First, we collected 1,672,178 tweets using 32 breast cancer-related keywords and the Twitter search API from May 27, 2018 to October 13, 2018. After filtering out duplicates and non-English tweets, 1,467,783 tweets were left. Second, we used the same list of keywords to identify relevant tweets from a database of random public tweets we collected from January 1, 2013 to December 30, 2017. We found 257,045 tweets from this database, within which 167,205 tweets were written in English. Due to the different mechanisms behind the two Twitter APIs (i.e., streaming vs. search), the volume of the tweets from the two data sources were significantly different. For Twitter search API, users can retrieve almost all public tweets related to the provided keywords within 10 to 14 days (i.e., the exact time range is not published by Twitter); while the Twitter streaming API returns a random sample (i.e., roughly 1% to 20% varying acros the years) of all public tweets at the time and covers a wide range of topics. The number of tweets related to breast cancer in the random sample database was expected to be low. We plot the trends of the tweet volumes for the two sources seperatly as shown in Figure 2.

*Table 1– Descriptive stateistics of the two data sources.*

| Data source | Data time range | # of tweets before pre-processing | # of Eng-lish tweets | # of ge-otagged tweets |
|---|---|---|---|---|
| Twitter search API | 05/27/2018-10/13/2018 | 1,672,178 | 1,467,783 | 428,041 |
| Random pub-lic tweets | 01/01/2013-12/30/2017 | 257,045 | 167,205 | 61,273 |

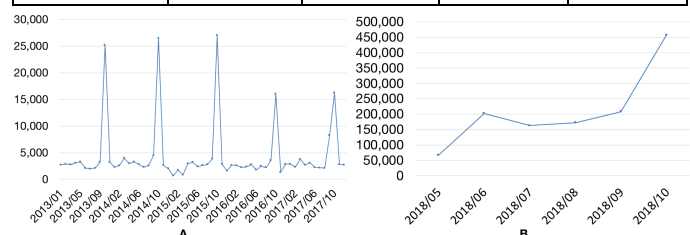

*Figure 2– Breast cancer tweet volume distributions across time (A: random public tweets; and B: Twitter search API).*

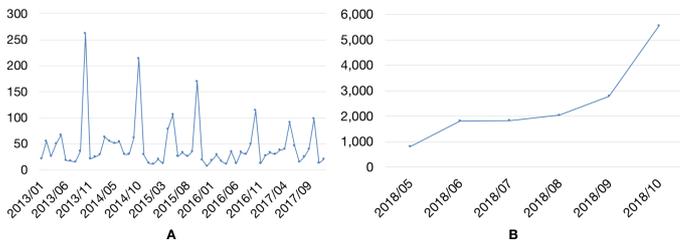

*Figure 3– The volume trends of tweets related to physical activity in laypeople's breast cancer discussions (A: random public tweets; B: Twitter search API).*

After integrating and eliminating duplicates and non-English tweets from the two sources, there were 1,634,988 unique tweets, in which 489,314 tweets can be geotagged to a US state.

To identify tweets related to PA from breast cancer tweets, a list of PA keywords (n=133) were developed through a snowball sampling process. Figure 3 shows the volume trends of tweets related to PA in laypeople's breast cancer discussions.

**Text classification**

We explored two deep learning classifiers to category the tweets. Both CNN and LSTM have been wildly used in text classifications and achieved state-of-the-art performance. Table 2 shows the performance of the different classifiers and tasks. We used 80% of the annotated data for training and the performance was measured on the rest 20% as an independent test data. As shown in Table 2, the CNN models outperformed the LSTM models in both tasks (i.e., 1) relevant vs. irrelevant; and 2) promotional information vs. laypeople's discussions). Thus, we adopted the CNN models as the final classifiers.

*Table 2– A comparison of classifier performance.*

| Classifier | Precision | Recall | F-score |
|---|---|---|---|
| **Task 1: Relevant vs. Irrelevant** | | | |
| CNN | 0.886 | 0.851 | 0.865 |
| LSTM | 0.847 | 0.797 | 0.814 |
| **Task 2: Promotional information vs. Laypeople's discussions** | | | |
| CNN | 0.943 | 0.937 | 0.941 |
| LSTM | 0.914 | 0.898 | 0.903 |

The CNN models identified 1,466,292 relevant tweets (out of 1,634,988 breast cancer related tweets). Out of the 1,466,292 relevant tweets, 961,110 are tweets with promotional information; and 505,182 tweets are laypeople's discussions.

**Sentiment analysis**

To answer RQ1 (i.e., "*How do people's emotions towards breast cancer change over time?*"), we assessed 5 emotion aspects (i.e., positive emotion, negative emotion, anxiety, anger, and sadness) of laypeople's breast cancer discussion tweets using LIWC. We visualized the changes of their emotion scores across time as shown in Figure 4. The Y-axis is the emotion scores generated by the LIWC. Even though there were fluctuations, the overall trends of negative emotion, anxiety, anger, and sadness have been decreasing since 2013 with significant Mann-Kendall test scores ($P_{trend} < .05$) as shown in Table 3.

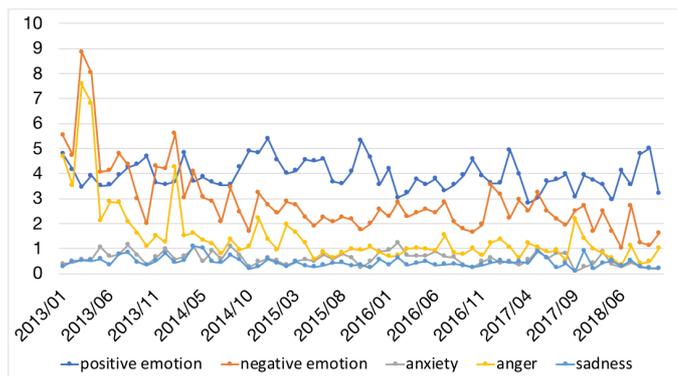

*Figure 4– Laypeople's emotion changes by time.*

*Table 3– Mann-Kendall tests of laypeople's emotion changes.*

| Emotion | P-value | Test-score | Trend |
|---|---|---|---|
| positive emotion | 0.17 | -1.37 | not significant |
| negative emotion | <0.01 | -5.45 | decreasing |
| anxiety | 0.01 | -2.53 | decreasing |
| anger | <0.01 | -5.44 | decreasing |
| sadness | 0.01 | -2.59 | decreasing |

We also compared the emotion scores of laypeople's breast cancer discussions across states, as heatmaps in Figure 5. The warmer the color the higher the emotion score. People in Mississippi had the highest negative emotion and anger when they discussed breast cancer on Twitter. Delaware had the highest positive emotion. Hawaii had the highest sadness. Washington D.C. had the highest anxiety.

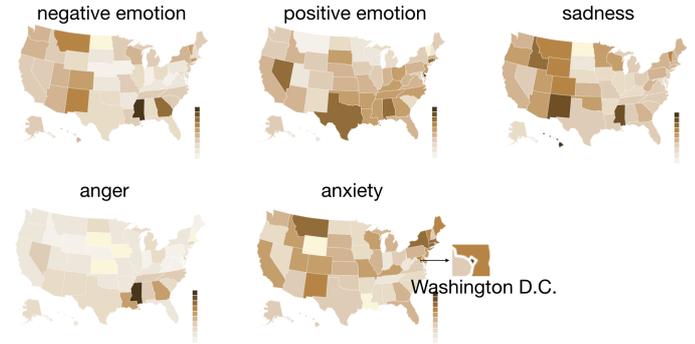

*Figure 5– Comparsion of laypeople's emotions towards breast cancer issues across states.*

We also analyzed laypeople's emotion changes over time by state. California (n=30,404) and Florida (n=17,983) had the highest number of breast cancer-related laypeople discussions, which gave us sufficient sample sizes to detect the trends. The trends of laypeople's emotion changes for these two states were decreasing as shown in Figure 6 and Table 4.

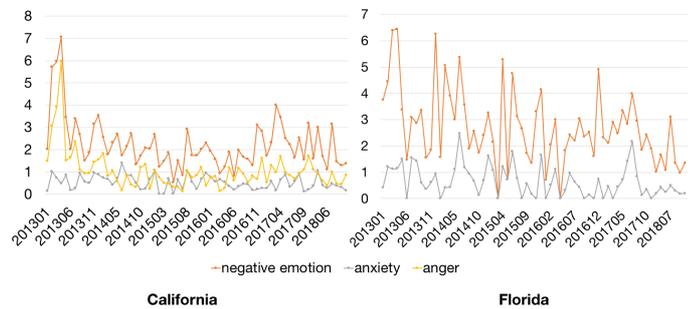

*Figure 6– Laypeople's emotion changes by time.*

*Table 4– Mann-Kendall test on laypeople's discussions from California and Florida*

| State | Emotion | P-value | Test-score | Trend |
|---|---|---|---|---|
| California | negative emotion | 0.01 | -2.47 | decreasing |
| | anxiety | 0.02 | -2.30 | decreasing |
| | anger | 0.03 | -2.16 | decreasing |
| Florida | negative emotion | <0.01 | -3.09 | decreasing |
| | anxiety | 0.01 | -2.60 | decreasing |

**Topic modeling**

To answer RQ 2-4, we used Biterm to discover the latent topics in our data. Based on experience from our previous work [7], we first set the number of topics to 100 to extract as many topics as possible. We then manually reviewed the 100 topics and a sample of assocaited tweets to assess topic quality and merge topics with similar themes. We summarized the 100 topics into 12 topics. The results are visualized as wordclouds in Figure 7.

The topic distributions of laypeople's discussions and promotional information are shown in Figure 8. Laypeople discussed more on the topics of "*family*", "*friend*", "*diagnosis*", "*food*", and "*treatment*"; while they talked less about "*risk*" and "*mortality*" compared with topics in promotional information.

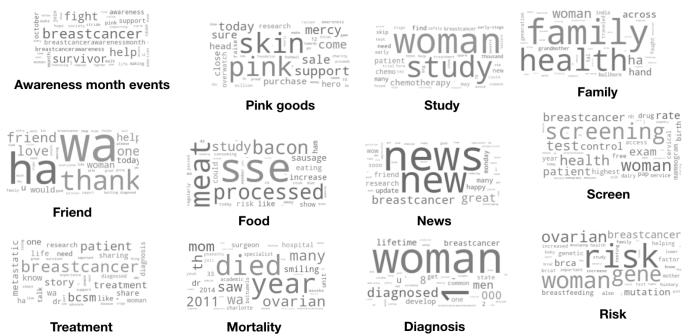

Figure 7– 12 topics summarized from the 100 topics learned with a Biterm topic model.

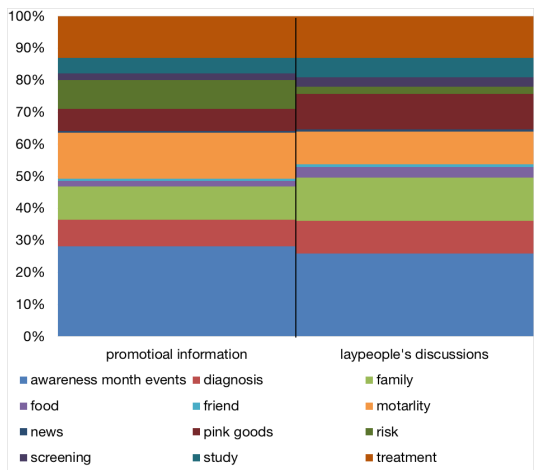

Figure 8– Topic distributions by tweet type (promotional information vs. laypeople's discussions).

To exam the relations between promotional information and laypeople's discussions by topic, we used the random tweet dataset (spanning from 2013 to 2017) and calculated the Pearson correlation coefficient between the two (promotional vs laypeople) based on their monthly tweet volumes for each topic. As shown in Table 5, laypeople's discussions had strong correlations with promotional information on "*awareness month events*", "*risk*", and "*treatment*" and moderate correlations on "*diagnosis*", "*family*", "*friend*", "*news*", "*pink goods*", "*screening*", and "*study*". The monthly trends from 2013 to 2017 for these topics are shown in Figure 9.

Table 5–Correlations between promotional information and laypeople's discussions based on tweet volumes by topic.

| Topic | Correlation coefficient | P-value |
|---|---|---|
| awareness month events | 0.711 | <0.001 |
| diagnosis | 0.635 | <0.001 |
| family | 0.682 | <0.001 |
| friend | 0.526 | <0.001 |
| news | 0.642 | <0.001 |
| pink goods | 0.557 | <0.001 |
| risk | 0.767 | <0.001 |
| screening | 0.690 | <0.001 |
| study | 0.683 | <0.001 |
| treatment | 0.788 | <0.001 |

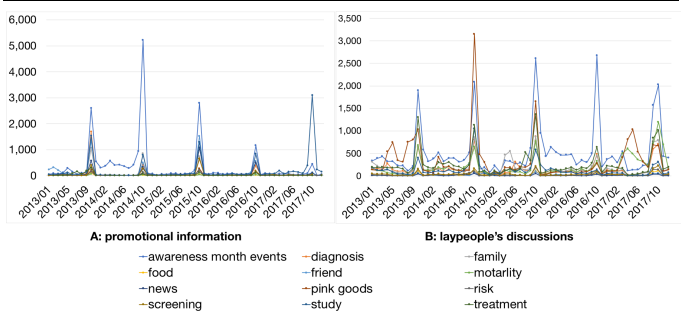

Figure 9– Selected topic trends from 2013 to 2017 that have significant correlations between promotional information and laypeople's discussiosns.

To answer RQ5, we first extracted all laypeople's breast cancer discussions that were aslo related to physical activites (PAs, i.e., discussions that contain PA-related keywords). We then extarcted 100 topics using the Biterm method on these PA-related dicssuions. Similar topics were merged into one; and two themes emerged: give support (i.e., give support to breast cancer awareness through sporting events) and reduce risk (i.e., raise awareness that PAs can reduce breast cancer risks). The wordclouds of the two themes are shown in Figure 10; and a few example tweets for each theme are shown in Table 5.

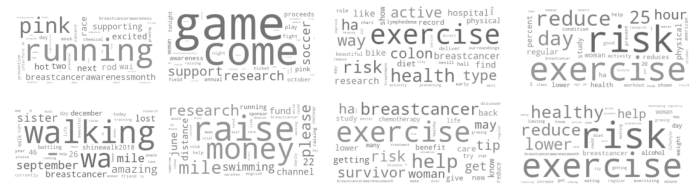

Figure 10– Two main themes in laypeople tweets related to both breast cancer and physical activities.

Table 5– Example of physcial activity related tweets by theme.

| Themes | Example tweets |
|---|---|
| A: give support to breast cancer awareness through sporting events | • *""I don't understand why it happened to me, but hope because I got cancer I can help bring about change." #BreastCancer survivor Shannon O'Fallon supports research with #Obliteride's 5K walk & urges others to help #curecancerfaster."*<br>• *"On May 14th of this year I lost my older sister Ashley to breast cancer. On August 11th my family will walk in her honor"* |
| B: physical activities can reduce the risk of breast cancer | • *"Important to think about HOW we #exercise, #breastcancer incidence and relapse risk can be reduced by physical activity https://t.co/ajfew2wY5D"*<br>• *"@RepDavidYoung I ran 91 miles in June. One of the reasons I run is bc it reduces my risk of recurrence of my breast cancer. O"* |

## Discussion and conclusions

The goals of our study were to examine breast cancer-related discussions on Twitter and to understand people's percetions and attitudes towards breast cancer through their Twitter posts. We were also interested in assessing how promotional information impacts laypeople's discussions on Twitter. Thus, we used well-established text analysis approaches (i.e., sentiment analysis and topic modeling) on breast cancer-related tweets to answer our five RQs.

As shown in our results (Figure 4), laypeople's attitudes towards breast cancer changed from time to time. The overall trends of negative affects (i.e., negative emotion, anger, anxiety, and sadness) have been decreasing since 2013. There have been some controversial issues being discussed on public news outlets related to breast cancer prevention and treatments, which might lead people think negatively. Taking the discussions on mammography as an example, mammography is a common way to screen for breast cancer as an early detection method. However, many people on Twitter questioned the effectiveness of mammography and raised concerns that it might bring overdiagnoses and overtreatments to patients. Nevertheless, as the negative attitudes are decreasing, it might indicate that stakeholders such as health organizations and agencies are doing a better job educating the public.

We also found that emotions in laypeople's breast cancer discussions on Twitter differ across the states. As shown in Figure 5, people in Mississippi had significantly higher negative emotion and anger when they discussed breast cancer-related issues on Twitter. Nevertheless, further investigations are needed to find the reasons behind these state differences. For example, people's attitudes towards breast cancer maybe a reflection of breast cacner screening rates. Mississippi is one of the states with the lowest breast cancer screening rate possibly due to barriers for people to access screening programs. These barriers may lead to negative emotions expressed in tweets (e.g., "*To cure a patient's disease at the cost of financial ruin falls short of our duty as physicians to serve*").

Laypeople's discussions were correlated with promotional information on a number of topics as shown in Table 5. These strong correlations might, from another perspective, indicate that breast cancer-related promotion strategies to raise public awareness have been rather successful in the past few years. Further, as observed in Figure 9, both promotional information and laypeople's discussions surged in October (i.e., breast cancer awareness month) every year, which suggested that promotional events in media (including social media) are effective ways to gain participants and raise public's awareness. Such raises in awareness would ultimately lead to improved health outcomes. For example, in early 1987, when the American Cancer Society started to focused on raising breast cancer awareness and before breast cancer became an official National Health Observances (NHO) event, only 26% of women in the U.S. had undergone a mammogram in the previous 12 months; while by October of the same year, the proportion had raised to 38% [12].

Over half of the women diagnosed with breast cancer gained weight during treatment with multiple reasons [13]. Chemotherapy often leads to fluid retention, reduced PA levels (due to pain and fatigue), decreased metabolism, and food cravings (that can reduce nausea). Nevertheless, breast cancer survivors should engage in a weight management program focusing on dietary intake and PA even during treatment if manageable. There is clear evidence that weight management including PAs have a positive impact on mental health during and after cancer treatment [14]. As shown in Figure 10, we obtained two general themes from laypeople's breast cancer discussions that are also related to PAs: 1) people like to give support (to breast cancer awareness through sporting events), and 2) people are aware of the benefits of PAs that can reduce breast cancer risks. When we looked into these tweets in more details, we also found tweets from breast cancer survivors (e.g., "*@raceforlife Fully intended to do this ( for the 5th year running; pardon the pun!!) but breast cancer diagnosis and treatment have run off with my fitness. Next year though!*"), which provides evidence data that social media might be a good source of information to study behavioral factors such as PAs related to improving surviorships including their quality of life (QoS). For example, we can further understand the barriers to adopting an exercise program among breast cancer survivors in our future work using social media data sources.

In sum, our study demonstrated that 1) social media such as Twitter are invaluable sources to provide insights into public and consumer health; and 2) natural language processing combined with machine learning are effective tools and methods to assess laypeople's attitudes changes in their health-related tweets, discover laypeople's perceptions towards specific health topics, and understand how promotional informaiton has an impact on laypeople's discussions.

## Acknowledgements

The study was supported by the University of Florida Health Cancer Center (UFHCC) Research Pilot Grant and in part by NSF Award #1734134 and NIH UL1TR001427. The content is solely the responsibility of the authors and does not necessarily represent the official views of the sponsors.

## Address for correspondence

Jiang Bian; Email: bianjiang@ufl.edu